\documentstyle[12pt]{article}

\begin{document}

\begin{titlepage}

\begin{center}
\baselineskip 24pt
{\Large {\bf Post-GZK Air Showers, FCNC, Strongly Interacting Neutrinos and
  Duality}}\\
\vspace{.5cm}
\baselineskip 16pt
{\large Jos\'e BORDES}\\
{\it Dept. Fisica Teorica, Univ. de Valencia,\\
  c. Dr. Moliner 50, E-46100 Burjassot (Valencia), Spain}\\
\vspace{.2cm}
{\large CHAN Hong-Mo, Jacqueline FARIDANI}\\
{\it Rutherford Appleton Laboratory,\\
  Chilton, Didcot, Oxon, OX11 0QX, United Kingdom}\\
\vspace{.2cm}
{\large Jakov PFAUDLER}\\
{\it Dept. of Physics, Theoretical Physics, University of Oxford,\\
  1 Keble Road, Oxford, OX1 3NP, United Kingdom}\\
\vspace{.2cm}
{\large TSOU Sheung Tsun}\\
{\it Mathematical Institute, University of Oxford,\\
  24-29 St. Giles', Oxford, OX1 3LB, United Kingdom}\\
\end{center}

\begin{abstract}
A review is given on the recently proposed idea that air showers with
energy $> 10^{20}$ eV beyond the Greisen-Zatsepin-Kuz'min cut-off may be
due to neutrinos having acquired a strong interaction at these energies,
as suggested by the Dualized Standard Model.  Such a hypothesis is shown
to be consistent with the so far known facts.  Further, by linking the 
astrophysical puzzle of post-GZK air showers through electric-magnetic 
duality to the problem of fermion generations in particle physics, one 
obtains on the one hand estimates for the rates of some flavour-changing 
neutral current decays which are accessible to experiments being planned, 
and on the other direct tests on the hypothesis performable by new air 
shower detectors such as Auger.  The suggestion does not exclude other
explanations for post-GZK showers given in the literature.  However, we
diagree with a recent paper by Burdman, Halzen and Gandhi which sweepingly 
claimed to have excluded nearly all explanations by new particle physics
including ours.

\end{abstract}

\end{titlepage} 

\clearpage

In what follows we shall review briefly an idea recently suggested in 
\cite{Bordesetal}.  It is a proposed marriage between two long-term puzzles 
in two rather different fields.  The first is the problem in cosmic ray 
physics of air showers with primary energy higher than $5 \times 10^{19}$ eV, 
which theoretically ought not to exist.  The other is in particle physics 
proper, namely the existence of 3 and only 3 generations of fermions, which 
up to now has had no generally accepted explanation.

Let me begin by outlining first the air shower puzzle.  Over the last 
thirty years, evidence has accumulated for the existence of air showers 
with primary energies greater than $10^{20}$ eV \cite{Volcano,Haverah,
Yakutsk,Flyseye,Agasa}.  To-date, about 9 such events have been recorded 
by several detectors using a variety of detection techniques, which makes 
it rather unlikely for all of them to be due to experimental biases or
errors.  These are dramatic events.  They occur some 12 km up in the
atmosphere, and when they hit the earth, they generally cover an area of
a few square kilometers in a shower containing as many as $10^{11}$ 
charged particles.  If we could see them with the naked eye, they would
be more spectacular than any fire-work display.  And according to Jakov
Pfaudler, the energy they carry is not far short of that of one of Boris
Becker's serves.  It is mainly to investigate further these so-called 
EHECR's (extremely high energy cosmic rays) that the huge Auger project 
is being planned, involving large arrays on two sites, one in each hemisphere, 
totalling in area 6000 km$^2$ \cite{Auger}.  

The reason for this unusual amount of interest, of course, is not because
they are spectacular but because they pose an intriguing question the answer
to which may reveal to us a new physics horizon.  The point is that air 
showers with such energies ought not in theory to be there at all.  Air 
showers at high energies are thought to be initiated mostly by protons, 
and protons at such an energy would quickly lose it by interacting with 
the photons in the 2.7 K microwave background field via, for example, the 
interaction:
\begin{equation}
p + \gamma_{2.7 K} = \Delta + \pi.
\label{pgammaint}
\end{equation}
Indeed, it has been shown by Greisen \cite{Greisen} and by Zatsepin and 
Kuz'min \cite{Zatmin} that the spectrum of protons originating from more 
than 50 Mpc away should be cut off sharply at around $5 \times 10^{19}$ eV 
in traversing the microwave background field. Thus, the observed air showers 
with energies in excess of $10^{20}$ eV at experimental energy resolutions 
of order 20 percent would be a blatant contradiction to theoretical 
expectations unless their primary protons have an origin within that sort
of distances.  However, such nearby sources are thought to be unlikely 
for the following reason.  Over such short distances, protons with these
extreme energies will be hardly deflected by the magnetic fields either
in our galaxy or in the space between.  The observed directions of the
air showers should thus point directly to their sources, but no candidate
sources have been found within a distance of 50 Mpc which are thought 
capable of producing particles of such an enormous energy.

A possible alternative explanation, among others, for these showers is that 
they are not initiated by protons at all but by some other particles.  Thus a 
stable zero-charged particle, such as a neutrino, could survive the long 
journey from whatever its extragalactic origin through the microwave background 
to arrive on earth with its high energy intact \cite{Sigljee,Halzen,Elmers}.
However, a neutrino with only the known electroweak interactions can readily 
penetrate our atmosphere.  In order to interact with the air at all to 
produce air showers at the observed frequency, cosmic neutrinos at these 
energies would need to have a very large flux, which is hardly imaginable.  
Even if this high flux is indeed available, the air showers induced would 
have angular and depth distributions which are at variance with those 
observed.  Whereas neutrino initiated showers are expected to be mostly 
horizontal, in order that the neutrino may pass through sufficient air for 
it to effect a collision, the observed events (with angular resolution of 
only $1^o - 2^o$), nearly all have incident angles of less than $40^o$ 
from the zenith.  Further, air showers initiated by weakly interacting 
neutrinos would have a flat distribution in the depth of atmosphere 
penetrated, not bunched at high altitudes as observed.  One concludes, 
therefore, that if neutrinos have only the known electroweak interactions, 
then the observed air showers with energies greater than $10^{20}$ eV are 
very unlikely to be initiated by neutrinos.

On the other hand, if neutrinos have interactions which become strong 
at ultra-high energies then the objections raised in the paragraph above
no longer apply and neutrinos may afford an explanation for the events under 
consideration.  This conjecture has been considered on the basis of possible 
substructures to quarks and leptons yet unknown to us \cite{Domokov,Domokos}.  
What we have suggested \cite{Bordesetal} is that there is another, perhaps
more attractive, theoretical scenario which will naturally give such 
interactions.  This leads then to the other puzzle we mentioned, namely
that of fermion generations in particle physics which goes as follows.

Neutrinos, like other leptons and quarks, are known to exist in three 
generations.  This fact has no explanation in the conventional
formulation of the standard model but is merely introduced into the theory 
as a phenomenological requirement, and as such has remained one of the 
greatest mysteries in particle physics.  Now, a favourite suggestion among 
theoreticians is that generations may in fact represent the quantum numbers 
of a broken continuous symmetry like $SU(3)$.  To bring it into line with 
other known continuous symmetries, we would then want this new one to be 
also a gauge symmetry.  If so it has to be mediated by a new set of gauge 
bosons and these, being flavoured but uncharged, would lead in turn to 
flavour-changing neutral currents (FCNC).  Now, such gauge bosons will have to 
be very heavy, for otherwise they will give rise to sizeable FCNC decays, which 
have not been observed.  Indeed, the strongest bounds on the gauge boson mass 
coming from $K$-decays are usually given to be in the 10 - 100 TeV region,
depending on the strength of the gauge coupling \cite{Cahnrari}.  If we accept 
this scenario, then at energies below, say, 100 TeV, generation-changing 
interactions due to the exchange of these bosons will be negligible and 
neutrinos will interact just via the usual electroweak forces.  However, 
at energies greater than 100 TeV, the new forces, which could in principle 
be strong, will come into play and give rise to new effects.

The incoming primary energy of the air showers under consideration is of
order $10^{20}$ eV, which in collision with a proton in the air corresponds
to a CM energy of around 400 TeV.  They are therefore, according to the
estimates of the preceding paragraph, at an energy possibly above the 
advent of the new interactions.  Hence, neutrinos at these energies may have 
already acquired strong interactions.  The first tenet then of our suggestion 
\cite{Bordesetal} is that they have and can therefore conceivably give rise 
to the post-GZK air showers observed.

Such a scenario, even if feasible, would of course still leave a number
of burning questions unanswered.  First, why should there be 3 and only
3 generations?  Second, why should the symmetry be a gauge symmetry and
why broken?  Third, why should the new interaction be strong?  And, fourth,
will the cross section of the neutrino with air nuclei be large enough 
then to produce air showers?  These questions cannot be answered in the
above general framework unless supplemented by more concrete assumptions.

A particular realization of this theoretical scenario capable of answering the 
questions raised is afforded by a recently proposed scheme \cite{Chantsou} based
on a nonabelian generalization of electric-magnetic duality \cite{Chanftsou}.  
In this scheme, the generation index is identified with dual colour, from 
which it follows that there are exactly three generations.  Further, it
follows from \cite{Chanftsou} that the generation symmetry is a gauge symmetry,
and from a well-known result of 't~Hooft's \cite{thooft} and the fact that
colour is confined, that the generation symmetry is broken.  Third, the 
broken generation symmetry is mediated by the dual gluons whose couplings 
are related to the usual couplings of colour gluons by the Dirac quantization 
condition, which in the standard conventions used in nonabelian theories 
with $\alpha = g^2/4\pi$ reads as \cite{Chantsou1}:
\begin{equation}
{\tilde g} g = 4 \pi,
\label{Diraccond}
\end{equation}
and are seen at these energies to be large, implying thus that the new 
interaction will be very strong.  As to the fourth question whether neutrinos 
will also acquire a large enough cross section with air nuclei to give the 
air showers observed, this is also answered in the affirmative but requires
a more detailed analysis to which we shall return later.  In addition to these
features, the scheme \cite{Chantsou} gives a CKM matrix which is the identity 
matrix at tree level but acquires mixing only from loop corrections, which 
within the scheme are amenable to calculation.  The values of the off-diagonal
CKM matrix elements in such a calculation will in general depend on the 
dual gluon mass which measures the onset energy scale of the new neutrino
interactions proposed.  Results from a recent calculation along these lines, 
the details of which we intend soon to report elsewhere \cite{Bordesetal1}, 
give a good fit to the experimental CKM matrix and are consistent with a 
dual gluon mass of around several 100 TeV.  Hence, in this scheme, not only 
is it possible for the neutrino to acquire strong interactions at energies 
above around 100 TeV as suggested in the general framework outlined above, 
but it seems that it is even {\it predicted} to be so.  If that is indeed 
the case, then air showers initiated by neutrinos with energies greater 
than $10^{20}$ eV would occur so long as neutrinos with such energies are 
produced somewhere out there in the universe.

Are there viable sources?  Since neutrinos are supposed to interact strongly
at such energies, then any source capable of accelerating protons to these
energies can produce neutrinos directly from collisions of the accelerated 
protons, a mechanism seemingly more efficient for high energy neutrinos 
than by, for example, pion decay.  Now, of the three possible candidate 
categories of sources lying above the line:
\begin{equation}
BR = E/Z,
\label{Hillas}
\end{equation}
on the Hillas plot \cite{Hillas} (where $B$ is the magnetic field in 
$\mu$G, $R$ the size in kpc, $E$ the energy in EeV = $10^{18}$ eV, and 
$Z = 1$ for protons), which are thought capable of accelerating protons to 
these energies, two are thought to have difficulty emitting them 
\cite{Boratav}.  For the neutron star, the accelerated proton is liable 
to lose its energy by sychrontron radiation on escaping simply by crossing 
the magnetic field which is itself responsible for its acceleration.  On 
the other hand, for active galactic nuclei, the accelerated proton is 
expected to suffer energy loss in its escape by interacting with the 
intense radiation field thought to surround the central parts of the AGN.  
We notice, however, that neither of these effects would affect the neutrino, 
which being neutral would not interact electromagnetically and would thus 
be able, once it is produced by the mechanism suggested above, to escape 
with its energy intact.

A neutrino interacting strongly at extreme energies would even offer
possible answers to several puzzling questions connected with the origin 
of $E > 10^{20}$ eV air showers.  For instance, three pairs among the observed
showers are known to have a common direction to within $2^o$ \cite{Hayashida}, 
suggesting thus a common origin for each pair.  However, if they are charged 
particles and have different energies as these pairs do, then they ought to be 
deflected differently by the intervening magnetic fields and arrive with
different directions unless the sources are rather close to earth.  This 
objection, however, does not apply to neutrinos so that each pair could
have come from the same distant source.  Further, it has been noted 
that the highest energy event known, namely the 320 EeV event recorded by
the Fly's Eye detector \cite{Flyseye}, points in the direction of a very
powerful Seyfert galaxy (MCG 8-11-11) which is 900 Mpc away \cite{Elmers}.  
If this is taken to be the source of that particular event, then one may 
wonder why such a source capable of producing a 320 EeV particle should 
give no signal in the 10 EeV range, which could be easily detected by
the Fly's Eye detector \cite{Boratav}.  This objection, however, poses 
no difficulty for the neutrino which interacts strongly only at extreme 
energies well above 100 TeV CM.  At lower energy, the interaction being 
there supposedly weak, neutrinos cannot, first of all, be produced directly 
from the collision of high energy protons as suggested above, and secondly, 
even if some of them are produced in MCG 8-11-11, the $\nu N$ cross 
section would have decreased sufficiently by these energies as to give 
them little chance of initiating air showers when they arrive on earth. 

Now, if such neutrinos are produced, by MCG 8-11-11 or some such object, then
they will be able to reach us.  They will be attenuated by neither the 2.7 K 
background photons since they are chargeless, nor by the 1.9 K background 
neutrinos, if massless, since their collisions will have CM energies of 
only around 200 MeV (even a neutrino with mass 10 eV will give only 40 GeV CM 
energy) at which the interaction is still very weak.  But, on their arrival on
earth, would they have sufficient cross sections with the air nuclei to 
produce showers with the observed properties, such as the above-mentioned 
angular and depth distributions?  Because a strong interaction, though 
necessary, is not sufficient to guarantee a large cross section.

The answer to this crucial question would seem to be yes if generation is
indeed dual colour as suggested in \cite{Chantsou}, but generally no if
generation-changing interactions are mediated by gauge bosons representing
an entirely new degree of freedom.  The reasoning goes as follows.  As is 
well-known, hadron cross sections are mainly governed by the sizes of the 
hadrons involved.  In ordinary $pp$ collisions, for example, one obtains 
a very reasonable estimate of around 100 mb for the total cross section 
if one simply pictures each proton as a greyish-black disc of radius around 
1 fermi.  Indeed, we know no better way than this for estimating the $pp$
total cross section, however sophisticated.  The reason that such a simple 
geometric picture works is because a parton in the incident proton, 
once it is inside the target proton, interacts with all partons of the 
target via long-ranged interactions so that it sees the target proton as 
a whole.  The situation is very different from that, say, of a neutrino at 
ordinary high energy interacting via only electroweak forces.  The range of 
the interaction being there given by the $W$ mass, the neutrino will see the 
proton only as a collection of grey dots representing the partons inside, 
each with radius $1/M_W$, giving thus much smaller cross sections.  Imagine
now what happens if the neutrino acquires a new interaction at ultra-high
energy via the exchange of some very high mass gauge bosons.  If these new
gauge bosons represent entirely new degrees of freedom, then the situation 
would be similar to the electroweak case, only now, because the interaction 
range is even shorter, the partons will appear as even smaller dots to the 
neutrino.  Assuming a greater strength for the coupling will not help since it 
will only change the grey dots into black dots, but cannot increase the cross 
section beyond the unitarity limit set by the size of the dots, or in other 
words the interaction range.  However, if one accepts that generation is 
dual colour as advocated in \cite{Chantsou}, then the situation is completely 
different.  The dual colour gauge bosons, as explained in \cite{Chantsou}, 
do not represent a different physical degree of freedom from the ordinary 
colour gluons but are related to the latter by an, unfortunately rather 
complicated, dual transform given in loop space \cite{Chanftsou}.  This 
fact was interpreted in physical terms in \cite{Chantsou} as a coupling 
between the dual and ordinary gauge bosons and allow the former to 
``metamorphose'' into the latter, so that on entering the target proton, 
the neutrino will interact at long range coherently with all partons in the 
target.  If so, it will see the target proton not as a collection of dots 
but as a disc, giving thus cross sections of hadronic size.  

Indeed, proceeding in this way from a geometric point of view, one can 
even give a rough  estimate of the cross section with air nuclei for 
ultra-high energy neutrinos \cite{Bordesetal} as follows.  Suppose that
the air nucleus does appear to the neutrino as a black disc of radius
$r_A$ but that the neutrino, with yet unknown internal structure, still
appears to the nucleus as a point.  Then the neutrino-nucleus cross
section is simply given by the area of the nuclear disc, namely $\pi r_A^2$.
Compare this now to the proton-nucleus cross section.  The nucleus will
still appear to the proton as a disc of radius $r_A$ but the proton now
will also appear to the nucleus as a disc of radius $r_p$.  If these
discs are black to each other, then a standard result of the geometric
picture gives the cross section as $\pi (r_p + r_A)^2$.  Further, assuming 
as often done that $r_A \sim r_p A^{1/3}$, $A$ being the atomic number of the 
air nucleus which we take on the average to be say 15, we obtain from this
that $r_A$ to be about 2.47 $r_p$.  From this we conclude that the
neutrino-nucleus cross section at the ultra-high energy we are interested 
in would be about half the proton-nucleus cross section at the same 
energy.  

Notice that in estimating the neutrino-nucleus cross section in the
geometric picture as we did above, we have not departed from the original
scheme in \cite{Chantsou} of ascribing the new generation-changing 
interaction to dual colour-exchange, nor have we been shirking our duty
in not trying to evaluate the cross section more properly.  The fact is
that hadron cross sections, involving as they do the coherent and strong
interaction between the constituents, is not available to perturbative
study, and we know of no better way than the geometric picture for dealing
with the problem.  Despite its crudeness, one beauty of the geometric picture
is that it is independent of much of the details of the inter-constituent
interactions, such as the coupling strength.  Nor is it dependent on
the energy except through the hadron size, which was already factored out
in giving the ratio of a half between the neutrino and the proton as we 
did above.

If the neutrino-nucleus cross section turns out to be indeed about a half
of the proton-nucleus energy, then it would be sufficient to produce air 
showers at the sort of depth as that observed in post-GZK events.  However,
the estimate should be regarded at best as rough and as only an upper limit 
in that, depending on the dual gluon mass, the dual colour interaction may 
not have attained full strength yet at these energies so that the nucleus 
may appear as grey rather than black to the neutrino.  Nevertheless, it 
will at least have a chance of being large enough to initiate air showers.

Suppose this is true.  Is there a way to subject the idea to further
experimental tests?  We can think of two ways for doing so in two rather
different directions.

First, as already explained above, exchanges of generation-carrying gauge
bosons will give rise to flavour-changing neutral current reactions which 
have not so far been observed and this nonobservation has been translated 
into a lower bound on the gauge boson mass.  Now, however, provided one 
accepts the hypothesis that the observed air showers with greater than 
$10^{20}$ eV energies are initiated by neutrinos then, independently of
the dual colour interpretation of \cite{Chantsou}, one would obtain an 
upper limit of the order of around 400 TeV for the mass of the mediating
gauge boson.  Together with the lower limit of around 100 TeV obtained 
from the bounds on FCNC K-decays, these would limit the gauge boson mass 
within sufficiently narrow limits to make predictions of FCNC decays in 
other reactions meaningful.  In Table 1, we list the branching ratios so 
predicted, assuming a unique gauge boson mass, for various FCNC decay modes 
of $s, c, b$ and $t$ particles which should be available for scrutiny at 
Daphne, BaBar and other strange-, charm-, bottom- and top-factory experiments 
now being planned.  Calculations along the lines of the dual scheme of 
\cite{Chantsou} will give more detailed predictions which are under
investigation \cite{Bordesetal3}.  The observation of FCNC decays can thus 
provide a test, though an indirect one, for the above suggested scenario.

\begin{table}

\begin{eqnarray*}
\begin{array}{||l|l|l||}
\hline \hline
  & Theoretical \, \, Estimate & Experimental \, \, Limit \\
Br(K^+ \rightarrow \pi^+ ll') & f_{s \rightarrow d l l'} 
\left(\frac{\tilde{g}^2}{4\pi}\right)^2 2 \times 10^{-12} &
2.1 \times 10^{-10}   \\
Br(K^0_s \rightarrow ll') & f_{sd \rightarrow l l'} 
\left(\frac{\tilde{g}^2}{4\pi}\right)^2 9 \times 10^{-11} &
3.2 \times 10^{-7}  \\
Br(D^+ \rightarrow \pi^+ ll') & f_{c \rightarrow u l l'} 
\left(\frac{\tilde{g}^2}{4\pi}\right)^2 2 \times 10^{-13} &
1.8 \times 10^{-5}  \\
Br(B^+ \rightarrow \pi^+ ll') & f_{b \rightarrow d l l'} 
\left(\frac{\tilde{g}^2}{4\pi}\right)^2 10^{-10} &
3.9 \times 10^{-3}  \\
Br(B^+ \rightarrow K^+ ll') & f_{b \rightarrow s l l'} 
\left(\frac{\tilde{g}^2}{4\pi}\right)^2 10^{-10} &
6 \times 10^{-5}  \\
\Gamma(t \rightarrow q ll') & f_{t \rightarrow q l l'} 
\left(\frac{\tilde{g}^2}{4\pi}\right)^2 9 \times 10^9 s^{-1} &
 \\
\hline \hline
\end{array}
\end{eqnarray*}

\caption{The estimates given above assume a unique mass of 400 TeV for the
gauge bosons with the gauge coupling ${\tilde g}$.  The coefficients
$f$ involve the mixing angles but are bounded by and of order unity.
For the dual scheme of \protect\cite{Chantsou}, ${\tilde g}$ is given by the
Dirac quantization condition (\protect\ref{Diraccond}) in terms of the ordinary
colour gluon coupling run to 400 TeV, and corresponds to a value for
$({\tilde g}^2/4\pi)^2$ of around 250.  The resulting branching ratios
satisfy the present experimental bounds but are accessible to new experiments
now being planned.  Detailed calculations with nondegenerate gauge boson
masses and explicit $f$'s depending on mixing angles will be reported 
elsewhere.}

\end{table}

Another test for the hypothesis, a direct one with air showers, is also 
available if one takes account of the estimate given above for the neutrino
cross section with air nuclei.  The cross section of a primary particle 
with the air nucleus governs the penetration depth of the shower it 
initiates.  Given the cross section of a particle, it is not difficult to 
calculate the distribution in penetration depth of the primary vertices
of the air showers it initiates.  For instance, inputting the proton-air
nucleus cross section of about 420 mb obtained from an extrapolation of
lower energy data, one easily obtains that the distribution of primary
vertices would peak at around 21 km in height for a vertical shower.
On the other hand, if we input a cross section of only half that size, say for 
the neutrino, the distribution would peak at only around 15 km in height.
These statements are only weakly dependent on the energy since the cross
sections on which they rely are also weakly dependent on energy.  Hence,
if it is really true that pre-GZK air showers are due mostly to protons 
and post-GZK air showers to neutrinos as we proposed, then there should
be a clear distinction in the depth-distribution of primary vertices for
the two categories, with the post-GZK showers clustering some 6 km lower
in height compared with the pre-GZK showers.  Unfortunately, in most
existing detectors, except perhaps for the Fly's Eye, the height of the
primary vertex of an air shower is not an easy quantity to determine.
We have therefore not yet been able to ascertain whether the above 
prediction is correct.  So far, we have found only one piece of information 
which may have a bearing on the matter, namely the so-called development
profile given by Fly's Eye for the highest energy shower ever recorded
at 3.2 $\times 10^{20}$ TeV.  Light for this shower began to be detected 
at an equivalent vertical height of around 12 km.  If one naively assumes
that this can be identified with the primary vertex, then the shower would 
seem much more likely to be a neutrino as we suggested than a proton, given
that the probability as calculated from the distribution of finding a 
proton primary vertex as low as 12 km in height is only about 5 percent
\cite{Bordesetal}.

Although neither of the tests suggested above can immediately be carried 
out, it would seem that with the new experiments being planned, for FCNC
decays those already mentioned and for air showers those large projects
such as Auger \cite{Auger}, one would have a chance to determine whether
the idea that post-GZK air showers are due to neutrino is likely to be
correct.

We note that although we believe that the explanation we have suggested 
for the post-GZK air showers is feasible, and even particularly interesting 
in that it is accessible to experimental tests both in air showers physics 
itself and in FCNC decays, we make no claim at all that it excludes other 
explanations.  Indeed, many other explanations have already been suggested,
both particle physical and astrophysical, ranging from `gluinos' in 
supersymmetry to cosmic topoplogical defects, and many appear possible, 
so that a much deeper study is necessary before we can decide on the right 
one.

However, we disagree with the conclusion of a paper which has recently 
appeared on the hep-ph bulletin board by Burdman, Halzen and Gandhi 
\cite{Burdzendhi} which claims sweepingly that no particle physics 
explanation with `scale' $>$ GeV is possible, excluding thus also ours 
presented above.  Their arguments centre on the question whether the cross 
section would be large enough to produce the air showers observed, involving 
thus only particle physics and no astrophysics.  In their explicit reference 
to our work, they based their conclusion only on a calculation of the 
first-order perturbative diagram exchanging a vector boson of large mass.  
In that case, it is obvious that the cross section of the neutrino will be 
much too small to explain post-GZK air showers, as already noted in our paper
\cite{Bordesetal} and outlined above.  For some reason, the effect of 
the `metamorphosis' of the dual gluon was completely ignored which, as 
explained above, was the basis of our estimate of a high energy $\nu A$-cross 
section about one-half of that of $p A$.  More generally, the claim in
\cite{Burdzendhi} of excluding almost all particle physics explanations
for post-GZK air showers appears also to be based just on first-order
perturbation theory and what was referred to `s-wave unitarity'.  It seems 
to us, however, that to estimate hadronic cross sections in general, it is 
imperative to take account of the coherent interaction of the hadron 
constituents (as explained e.g. in \cite{Bordesetal}), which interaction 
is essentially nonperturbative and typically involves many partial waves
\cite{Weinberg}.  It is therefore, to us, not at all surprising that an 
estimate, as that in \cite{Burdzendhi}, based only on s-wave unitarity 
and first-order perturbative calculation of the interaction between 
individual constituents, is unable to give a correct result.

\vspace{.5cm}

\noindent {\large {\bf Acknowledgement}}

\vspace{.2cm}

We thank Jeremy Lloyd-Evans for first interesting us in high energy air
showers. Besides, one of us (JB) acknowledges support from the Spanish 
Government on contract no. CICYT AEN 97-1718, while another (JP) is 
grateful to the Studienstiftun d.d. Volkes and the Burton Senior 
Scholarship of Oriel College, Oxford for financial support.

\end{document}